\documentclass[prc,aps,showpacs,twocolumn,floatfix]{revtex4}
\usepackage{graphicx}
\begin{document}
\title{Family of Hermitian Low-Momentum\\ 
Nucleon Interactions with Phase Shift Equivalence}

\author{ Jason D. Holt $^{1}$, T. T. S. Kuo$^{1}$, G.E. Brown $^{1}$}
\affiliation{ $^1$Department of Physics, SUNY, Stony Brook, New York 11794, USA}
\date{\today}

\begin{abstract}
Using a Schmidt orthogonalization transformation, a family of
Hermitian low-momentum NN interactions is derived from the non-Hermitian
Lee-Suzuki (LS) low-momentum NN interaction. 
As  special  cases, our transformation reproduces the Hermitian interactions
of Okubo and Andreozzi. Aside from their common preservation of the deuteron
binding energy, these Hermitian interactions are shown to be phase shift
equivalent, all preserving the empirical phase shifts up to 
decimation scale $\Lambda$.
 Employing a solvable matrix model, the Hermitian
interactions given by different orthogonalization transformations 
are studied; the interactions can be very different from each other 
particularly when there is a strong intruder state influence. 
However, because the parent LS low-momentum NN interaction is 
only slightly non-Hermitian,
the Hermitian low momentum nucleon interactions given by our
transformations, including
the Okubo and Andreozzi ones, are all rather similar to each other.
Shell model matrix elements given by the
 LS and several Hermitian low momentum interactions are compared.

\end{abstract}

\pacs{21.60.Cs}
\maketitle

\newpage

\section{Introduction}
A fundamental problem in nuclear physics has been the determination
of the effective nucleon-nucleon ($NN$) interaction appropriate
for complex nuclei. Typically, one starts from a NN interaction, $V_{NN}$,
constrained by the deuteron properties and the empirical 
low-energy NN scattering phase shifts. Several realistic 
meson models \cite{cdbonn,nijmegen,argonne,paris} for $V_{NN}$
have been obtained in this way, and while they all share the same one pion 
tail, they differ significantly in how they treat the shorter distance
components. Despite this difference, these models all give approximately
the same {\bf low-energy} phase shifts and deuteron binding energy.  
This result clearly manifests the main theme
of the renormalization group (RG) and effective field theory (EFT) 
approach, namely  physics in the infrared region is insensitive to the 
details of the short-distance dynamics
\cite{meissner,lepage,kaplan98,epel99,EFT,EFT02,EFT03,kolck,haxton,beane}.
It is thus possible to have infinitely many theories that differ 
substantially 
at small distances, but still give the same low-energy physics, as long as
they possess the same symmetries and the ``correct'' long-wavelength 
structure. Since low-energy physics isn't concerned with these high-energy
details, one should just use an effective theory
with the short-wavelength modes integrated out.

Following this RG-EFT idea, a low-momentum NN effective 
interaction, $V_{low-k}$, was recently developed 
\cite{bogner01,bogner02,kuorg02,schwenk02,coraggio02,coraggio02b}.
While similar in spirit to traditional EFT, $V_{low-k}$ is not derived via
the usual RG-EFT methods; rather, it combines the standard nuclear 
physics approach (SNPA) with EFT making it a {\it more effective} 
EFT (MEEFT) \cite{brownrho,brownrho02,park}, as discussed in \cite{kubodera}.
 As such, a main step in its derivation is
the integrating out of the high momentum components 
of some realistic NN potential model $V_{NN}$ such as those of Ref.
\cite{cdbonn,nijmegen,argonne,paris}. Even though these $V_{NN}$ models
are quite different from each other, it is remarkable that 
the $V_{low-k}$'s derived from them
are nearly identical to each other, suggesting a nearly unique $V_{low-k}$
\cite{bogner03}.
Furthermore, shell model calculations using $V_{low-k}$ have given
very encouraging results over a wide range of nuclei
\cite{bogner02,kuorg02,coraggio02,coraggio02b}. Applications of $V_{low-k}$
to quasi-particle interaction, superfluid gaps and equation of state
for neutron matter have also been highly successful \cite{schwenk02}.

An important problem in deriving $V_{low-k}$ is how to obtain a 
low-momentum NN interaction which is Hermitian. The $V_{low-k}$
given by the T-matrix equivalence approach
\cite{bogner01,bogner02} is not Hermitian, and some additional transformation
is needed to make it Hermitian.  There are a number of
methods used to obtain a Hermitian effective interaction, such as those of
Okubo \cite{okubo}, Suzuki and Okamoto \cite{suzuki83}, and
Andreozzi \cite{andre96}. Which of these methods should one use?
How different are the Hermitian $V_{low-k}$'s  given by them?
These questions seem to have not been investigated.
In the present work, we shall study the $V_{low-k}$'s given by 
these methods as well as develop a unified method by which
a family of phase-shift equivalent Hermitian low-momentum NN interactions
can be obtained. It is, of course, important that $V_{low-k}$ 
preserve phase shifts, and while the non-Hermitian $V_{low-k}$ given by 
the T-matrix equivalence \cite{bogner01,bogner02} can be shown
to preserve phase shifts in a straightforward way, 
it seems to be more involved to prove the phase shift preservation
by Hermitian $V_{low-k}$. Epelbaoum et al. \cite{epel99}
have pointed out that phase shifts are preserved for the Okubo low-momentum
interaction \cite{epel99}, but phase shift preservation for other Hermitian
$V_{low-k}$'s seems to have not been investigated - an issue we will also
examine.

To conclude our introduction, we shall briefly review the non-Hermitian
low-momentum interaction given by the T-matrix equivalence approach.
Then in Section II we present a general method, based on
Schmidt orthogonalization
transformation, for generating a family of Hermitian effective interactions,
and show that the Hermitian interactions of Okubo, Suzuki and
Okamoto and Andreozzi all belong to this family.
In Section III, we  study our method using a 
simple solvable matrix model of the Hoffmann type \cite{hoff}; focusing
on the influence of intruder states and the difference of the 
Hermitian interactions given by the various methods. 
In Section IV we present a proof that phase shifts are preserved
by the Hermitian interactions generated by our method. 
This preservation will also be checked by phase shift calculations
using different Hermitian interactions.
Finally in Section V we present our results for the $V_{low-k}$'s 
corresponding to  the
various Hermitian interactions, where we show that our approach allows us to
construct a $V_{low-k}$ which preserves the deuteron wave function,
in addition to the preservation of deuteron binding energy and
low-energy phase shifts.

In the following let us first briefly review the $V_{low-k}$ given by
T-matrix equivalence \cite{bogner01,bogner02}.
 One starts from the half-on-shell $T$ matrix
\begin{eqnarray}
  T(k',k,k^2)  = V_{NN}(k',k)   +P \int _0 ^{\infty} q^2 dq  V_{NN}(k',q) 
\nonumber \\
  \times \frac{1}{k^2-q^2 +i0^+ } T(q,k,k^2 ) .
\end{eqnarray}
An effective low-momentum $T$ matrix is then defined by 
\begin{eqnarray}
\lefteqn{T_{low-k }(p',p,p^2)  = V_{low-k }(p',p)} \\
&&  + P\int _0 ^{\Lambda} q^2 dq  V_{low-k }(p',q) 
 \frac{1}{p^2-q^2 +i0^+ } T_{low-k} (q,p,p^2) \nonumber 
\end{eqnarray}
where the intermediate state momentum is integrated up to $\Lambda$.
In the above two equations we employ the principal value boundary
conditions, as indicated by the symbol $P$ in front of the integral
sign.  We require the $T$ matrices satisfy the condition
\begin{equation}
 T(p',p,p^2 ) = T_{low-k }(p',p, p^2 ) ;~( p',p) \leq \Lambda .
\end{equation}
The above equations define the effective low-momentum interaction
 $V_{low-k}$. Using a $\hat Q$-box folded-diagram method 
\cite{klr71,ko90}, it has been shown \cite{bogner01,bogner02,kuorg02}
that the above equations are satisfied by the solution  
\begin{eqnarray}
  V_{low-k} &= & \hat{Q} - \hat{Q'} \int \hat{Q} + \hat{Q'} \int \hat{Q} \int 
  \hat{Q}  \nonumber \\
   && {} - \hat{Q'} \int \hat{Q} \int \hat{Q} \int \hat{Q} + \cdots ,
\end{eqnarray} 
where $\hat Q$-box denotes the irreducible vertex function whose 
intermediate states are all beyond $\Lambda$, and $\hat Q'$ is the same 
vertex function except that it starts with terms second order in the 
interaction. The low-momentum effective NN interaction of Eq.(4) can be 
calculated using iteration methods such as the Lee-Suzuki \cite{suzuki80},
Andreozzi \cite{andre96} or Krenciglowa-Kuo \cite{krmku74} methods. 

The above $V_{low-k}$ preserves both the deuteron binding energy and the 
half-on-shell T-matrix of $V_{NN}$ (which implies the preservation of the
phase shifts up to $E_{lab}=2\hbar^2\Lambda^2 /M$, $M$ being the nucleon mass). 
This $V_{low-k}$ is not Hermitian, as indicated by Eq.(4).   
As we will show soon, starting from this $V_{low-k}$
a family of  phase-shift equivalent
Hermitian low-momentum NN interactions can be obtained.

\section{Formalism}

Before presenting our general Hermitization procedure, let us first review 
some basic formulations about the model space effective
interaction. We start from the Schroedinger equation
\begin{equation}
(H_0+V)\Psi _n =E_n \Psi _n,
\end{equation}
where $H_0$ is the unperturbed Hamiltonian and V the interaction. The 
eigenstates of $H_0$ are $\phi _n$ with eigenvalues
$\epsilon _n$. For example, $H_0$ can be the kinetic energy operator and V 
the NN interaction $V_{NN}$. A model-space
projection operator $P$ is defined as $\sum _{i=1}^d \mid \phi _i \rangle
\langle \phi _i \mid$, where d is the dimension of the model space.
The projection operator complement to P is denoted as Q, and
as usual, one has $P^2=P,Q^2=Q$ and $PQ=0$.
In the present work, $P$ represents all the momentum states with momentum
less than the cut-off scale $\Lambda$.

 A model-space effective interaction $V_{eff}$ is introduced with the 
requirement that the effective Hamiltonian $P(H_0+V_{eff})P$ reproduces
some of the eigenvalues and certain information about the eigenfunctions
of the original Hamiltonian $(H_0+V)$. There are a number of ways to derive
$V_{eff}$, but, as indicated by Eq.(4), our effective interaction is
obtained by the folded diagram method \cite{klr71,ko90} and can be 
calculated conveniently using the Lee-Suzuki-Andreozzi \cite{andre96}
 or Krenciglowa-Kuo \cite{krmku74}
iteration methods. We denote this effective interaction
as $V_{LS}$, with the corresponding model space Schroedinger equation
\begin{equation}
P(H_0+V_{LS})P\chi _m =E_m \chi _m,
\end{equation}
where $\{E_m\}$ is a subset of $\{E_n\}$ of Eq.(5) and
$ \chi _m =P \Psi _m $.

It is convenient to rewrite the above effective interaction in terms
of the wave operator $\omega$, namely
\begin{equation}
PV_{LS}P =Pe^{-\omega}(H_0 +V)e^{\omega}P-PH_0P,
\end{equation}
where $\omega$ possesses the usual properties
\begin{eqnarray}
\omega &=& Q\omega P; \nonumber \\
\chi_m &=& e^{-\omega} \Psi _m ; \nonumber \\
\omega \chi_m &=& Q\Psi _m.
\end{eqnarray}

While the eigenvectors $\Psi _n$ of Eq.(5) are orthogonal to each other, it is 
clear that the eigenvectors $\chi _m$ of Eq.(6) are not so and the 
effective interaction $V_{LS}$ is not Hermitian. We now make a 
$Z$ transformation such that 
\begin{eqnarray}
Z \chi_m &=& v _m; \nonumber \\
\langle v_m \mid v_{m'} \rangle &=& \delta _{mm'};~m,m'=1,d,
\end{eqnarray}
where d is the dimension of the model space. This transformation
reorients the vectors $\chi_m$ such that they become orthonormal
to each other.
We assume that $\chi _m$'s ($m=1,d$) are linearly independent so that
$Z^{-1}$ exists, otherwise the above transformation is not possible.  
Since $v_m$ and $Z$ exist entirely within the model space, we can write
$v_m=Pv_m$ and $Z=PZP$. 

Using Eq.(9), we transform Eq.(6) into
\begin{equation}
Z(H_0+V_{LS})Z^{-1}v_m=E_m v_m,
\end{equation}
which implies
\begin{equation}
Z(H_0+V_{LS})Z^{-1}=\sum _{m\epsilon P} E_m \mid v_m \rangle 
\langle v_m \mid.
\end{equation}
Since $E_m$ is real (it is an eigenvalue of Eq.(5)) and the vectors $v_m$
are orthonormal to each other, 
$Z(H_0+V_{LS})Z^{-1}$ must be Hermitian.  
The original problem is now reduced to a Hermitian model-space 
eigenvalue problem
\begin{equation}
P(H_0 +V_{herm})Pv_m=E_m v_m
\end{equation}
with the Hermitian  effective interaction 
\begin{equation}
V_{herm}=Z(H_0+V_{LS})Z^{-1}-PH_0P,
\end{equation}
or equivalently
\begin{equation}
V_{herm}=Ze^{-\omega}(H_0+V)e^{\omega}Z^{-1}-PH_0P.
\end{equation}


To calculate $V_{herm}$, we must first have the $Z$ transformation. 
Since there are certainly many ways to construct $Z$, this generates
a family of Hermitian effective interactions, all originating from $V_{LS}$.
For example, we can construct Z using the familiar Schmidt 
orthogonalization procedure, namely:
\begin{eqnarray}
v_1&=&Z_{11}\chi _1  \nonumber  \\
v_2&=&Z_{21}\chi _1 +Z_{22} \chi _2  \nonumber  \\
v_3&=&Z_{31}\chi _1 +Z_{32} \chi _2  +Z_{33} \chi _3 \nonumber  \\
v_4&=&......,
\end{eqnarray}
with the matrix elements $Z_{ij}$ determined from Eq.(9). 
We denote the Hermitian effective interaction using this Z 
transformation as $V_{schm}$.  Clearly there are 
more than one such Schmidt procedures. For instance, we can 
use $v_2$ as the starting point, which gives
$v_{2} = Z_{22}\chi _2$, $v_3=Z_{31} \chi _1+Z_{32}\chi_2$,
and so forth. This freedom in how the orthogonalization is
actually achieved, gives us infinitely many ways to generate a Hermitian 
interaction, and this is our family of Hermitian interactions produced from
$V_{LS}$.

We now show how some well-known Hermitization transformations relate to 
(and in fact, are special cases of) ours.
We first look at the Okubo transformation \cite{okubo}. From Eq.(8) we have
\begin{equation}
\langle \chi _m \mid (1+\omega ^+ \omega )\mid \chi _{m'} \rangle
=\delta _{mm'}.
\end{equation}
It follows that an analytic choice for the $Z$ transformation is
\begin{equation}
Z=P (1+\omega ^+ \omega)^{1/2}P.
\end{equation}
This leads to the Hermitian effective interaction
\begin{eqnarray}
V_{okb1}&=&P (1+\omega ^+ \omega)^{1/2}P(H_0+V_{LS})P 
 (1+\omega ^+ \omega)^{-1/2}P \nonumber \\
  & & -PH_0P.
\end{eqnarray}
From Eqs. (8), (9), (16) and (17), it is easily seen that the above is
equal to the Okubo Hermitian effective interaction
\begin{eqnarray}
V_{okb}&=& P (1+\omega ^+ \omega)^{-1/2}(1+\omega^+)(H_0+V) \nonumber \\
& & \times (1+\omega) (1+\omega ^+ \omega)^{-1/2}P -PH_0P
\end{eqnarray}
giving us an alternate expression, Eq.(18), for the Okubo interaction.

There is another interesting choice for the transformation Z. As pointed 
out by Andreozzi \cite{andre96}, the positive definite operator
$P(1+\omega ^+ \omega)P$ can be decomposed into two Cholesky matrices,
namely
\begin{equation}
P (1+\omega ^+ \omega)P= PL L^TP,
\end{equation}
where $L$ is a lower triangle Cholesky matrix, $L^T$ being its transpose.
Since $L$ is real and it is within the P-space, we have from Eq.(16) that 
\begin{equation} 
Z=L^T 
\end{equation}
and the corresponding Hermitian effective interaction from Eq.(13) is
\begin{equation}
V_{cho}=PL^TP(H_0+V_{LS})P(L^{-1})^TP -PH_0P.
\end{equation}
This is the Hermitian effective interaction of Andreozzi \cite{andre96}.

\begin{figure}
\rotatebox{270}{\scalebox{0.4}{\includegraphics{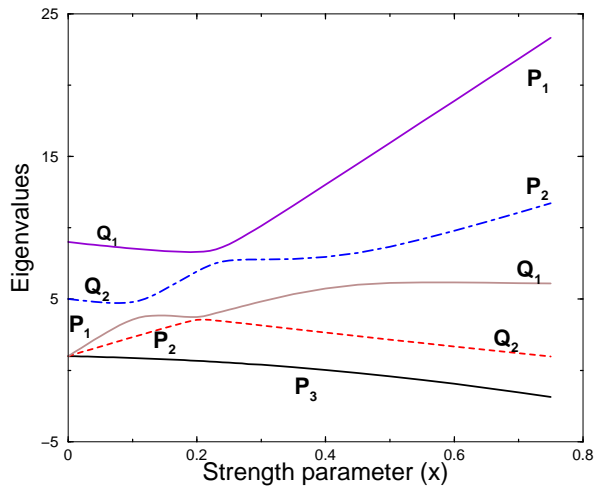}}}
\caption{Eigenvalues of our model $5 
\times 5$ interaction as a function of the strength parameter, $x$.  }
\label{fig.1}
\end{figure}

The final Hermitian effective interaction we consider is that of Suzuki 
and Okamota 
\cite{suzuki83,suzuki94}, which is of the form
\begin{equation}
V_{suzu}=Pe^{-G}(H_0+V)e^{G}P-PH_0P
\end{equation}
with $G=tanh^{-1}(\omega-\omega^{\dagger})$ and $G^{\dagger}=-G$.
It has been shown that this interaction is the same as the Okubo interaction 
\cite{suzuki83}.
In terms of the $Z$ transformation, it is readily seen that 
the operator $e^{-G}$ in Eq. (23) is equal to $Ze^{-\omega}$ 
with $Z$ given by Eq. (17). Thus, three well-known and particularly useful
Hermitian effective interactions indeed belong to our family.

\section{Model calculations}

\begin{table}
\caption{Comparison of 3x3 Hermitian ($okb,cho,schm$) and non-Hermitian
($LS$) effective interactions. The matrix model of Eq.(24) is employed
with interaction strength x=0.1.}
\vskip 0.5cm
\begin{tabular}{|ccc|c|}
\hline
-0.0368 & 0.3657 & 0.3976 & $V_{LS}$        \\
 0.3793 & 0.8811 & 0.4482 &                  \\
 0.4722 & 0.4728 & 1.3999 &                  \\ \hline

-0.0358 & 0.3732 & 0.4368 & $V_{okb}$     \\
 0.3732 & 0.8814 & 0.4609 &                  \\
 0.4368 & 0.4609 & 1.3986 &                  \\ \hline

-0.0041 & 0.4063 & 0.5215 & $V_{cho}$  \\
 0.4063 & 0.9020 & 0.5143 &                  \\
 0.5215 & 0.5143 & 1.3461 &                  \\ \hline

-0.0251 & 0.3823 & 0.4698 & $V_{schm}$   \\ 
 0.3823 & 0.8846 & 0.4695 &                  \\
 0.4698 & 0.4695 & 1.3846 & \\ \hline

\end{tabular}
\end{table}

The results from the previous section show that a family of Hermitian
effective interactions can be derived from a Schmidt-type transformation
of the non-Hermitian interaction $V_{LS}$. We now check to see if these
interactions do reproduce some of the eigenvalues of the
original Hamiltonian, and how the effective 
interactions given by the various methods differ.
In this section, we shall use a solvable matrix model to study
these questions.  Hoffmann et al. \cite{hoff} have employed a matrix
model to study the influence of intruder states on effective
interactions. Since we are also interested to see 
how intruder states might effect
our Hermitian potentials, we use a matrix model of this type to study
$V_{schm}$, $V_{okb}$ and $V_{cho}$,
together with their parent non-Hermitian interaction $V_{LS}$.

We employ a 5-by-5 matrix model $H=H_0 +x  V$, $x$ being a 
strength parameter. We take $H_0=\{1,1,1,5,9\}$ and 
\begin{equation}
V= \bordermatrix{
&  &   &  &  &   \cr 
&1 &5  &5 &0 &5  \cr
&5 &25 &5 &5 &0  \cr
&5 &5  &15 &2 &2  \cr
&0 &5  &2 &-5 &1  \cr
&5 &0  &2 &1  &-5  \cr  }.
\end{equation}
\vskip 0.4cm
\noindent Our P-space is chosen as the space spanned 
by the three lowest eigenstates
of $H_0$, namely $PH_0P=\{1,1,1\}$. The rather large diagonal matrix
elements are used for $V$ so that intruder states will enter
as the strength parameter $x$ increases.  How this happens is shown in 
Fig. 1, where the eigenvalues of H are plotted as a function of the strength 
parameter. In this figure the states are labelled
as $P_1$, $P_2$, $P_3$, $Q_1$ and $Q_2$, according to the structure
of their wave functions. 
The $P$ states are those whose wave functions are dominated by their
P-space components, i.e. $\langle \Psi | P|\Psi \rangle$. In contrast,
the $Q$ states are dominated by their Q-space components.
For a weak interaction (small x), the lowest three states are 
all $P$ states. 
As the interaction strength increases, states $Q_1$ and $Q_2$ decrease in
value and
states $P_1$ and $P_2$ increase, and we would expect these eigenvalues to 
intersect. Of course, they don't actually cross, but at certain interaction
strength a Q-state ``intrudes'' into the P-space, 
becoming lower than  the rising P-state.
At x=0.5, for example, the lowest three
states are $P_3$, $Q_2$ and $Q_1$, so $Q_2$ and $Q_1$ are intruder states 
in the sense that
they have entered the P-space when the interaction is strong.
We want our model-space effective Hamiltonian $PH_{eff}P$ to reproduce
the lowest three states of $H$. Thus at large x, we are requiring 
our $PH_{eff}P$ to reproduce two intruder states.  In the rest of this 
section we study how these intruder states effect our potentials and what 
impact they have on the Hermitization procedure.

\begin{table}
\caption{Comparison of 3x3 Hermitian ($okb,cho,schm$) and non-Hermitian
($LS$) effective interactions. The matrix model of Eq.(24) is employed
with interaction strength x=0.55.}
\vskip 0.5cm
\begin{tabular}{|ccc|c|}
\hline
-2.6246 & -3.4921 & -0.9479 & $V_{LS}$         \\
 0.8528 &  1.1730 &  2.7893 &                   \\
 0.2651 & 0.7817  & -3.0008 &                   \\ \hline

-1.7918 & -0.6049 &  0.0615 & $V_{okb}$      \\
-0.6049 &  0.6264 &  0.7647 &                   \\
 0.0615 &  0.7647 & -3.2870 &                   \\ \hline

-1.2079 &  0.4599 &  0.6813 & $V_{cho}$   \\
 0.4599 & -0.4459 &  1.3268 &                   \\
 0.6813 &  1.3268 & -2.7985 &                   \\ \hline

-1.9123 &  0.3729 &  0.4656 & $V_{schm}$    \\
 0.3729 & -0.5113 &  1.1263 &                   \\    
 0.4656 &  1.1263 & -2.0288 &  \\ \hline

\end{tabular}
\end{table}

The effective interactions are calculated using the following procedures. 
First we calculate the wave operator $\omega$ using the Lee-Suzuki 
iteration method developed by Andreozzi \cite{andre96}; we denote this 
method as the ALS method, and $V_{LS}$ is then given by Eq.(7). $V_{schm}$ is 
obtained from Eqs.(9), (13) and (15), and $V_{cho}$ is calculated
using Eqs.(20) and (22). For the Okubo interaction,
it is convenient to calculate it using the method of Suzuki et al.
\cite{suzuki83,suzuki94,kuo93}.
With this method, we first find the eigenvalues and eigenfunctions defined by
\begin{equation}
(1+\omega ^+ \omega)\mid \alpha \rangle = \mu _{\alpha} ^2
\mid \alpha \rangle.
\end{equation}
Then the Okubo Hermitian effective interaction is given by 
\begin{eqnarray}
\lefteqn{\langle\alpha|V_{okb}|\beta\rangle = D(\alpha,\beta)}  \\
&& \times \left(\sqrt{\mu^2_{\alpha}+1}\langle
\alpha|
V_{LS}|\beta\rangle+\sqrt{\mu^2_{\beta}+1}\langle\alpha|V_{LS}^{\dagger}|
\beta\rangle\right)
    \nonumber 
\end{eqnarray}
with 
\begin{equation}
D(\alpha,\beta)=\left[\sqrt{\mu_{\alpha}^2+1}
+\sqrt{\mu_{\beta}^2+1}\right]^{-1}.
\end{equation}

\begin{table}
\caption{Comparison of eigenenergies and wavefunctions for the Hermitian
($okb,cho,schm$) and non-Hermitian ($LS$) effective interactions.
 The matrix model of Eq.(24) is employed with
interaction strength x=0.55, where its 5 eigenvalues are -0.6714, 1.9146,
6.1702, 9.2084, and 17.4281.}
\vskip 0.5cm
\begin{tabular}{p{1in}cc}

                     & Eigenenergy &    Wavefunction          \\ [.25cm]
\cline{2-3}
  $V_{LS}$           & -0.6714     & (-0.959,  0.193,  0.207) \\
     $V_{okb} $ & -0.6714     & ( 0.881, -0.466, -0.081) \\
  $V_{cho} $ & -0.6714     & (-0.528,  0.790,  0.311) \\
   $V_{schm}$ & -0.6714     & (-0.959,  0.193,  0.207) \\ [.25cm]
\cline{2-3}
                     &  1.9146     & ( 0.916,  0.397, -0.051) \\
                     &  1.9146     & ( 0.472,  0.853,  0.223) \\
                     &  1.9146     & (-0.826, -0.563,  0.030) \\
                     &  1.9146     & ( 0.235,  0.951,  0.201) \\  [.25cm]  
\cline{2-3}
                     &  6.1702     & (-0.325, -0.179,  0.929) \\
                     &  6.1702     & (-0.195,  0.932,  0.306) \\
                     &  6.1702     & ( 0.198, -0.241,  0.950) \\
                     &  6.1702     & (-0.035, -0.234,  0.971) 

\end{tabular}
\end{table}

Using these methods we can calculate $V_{LS}$, $V_{okb}$, $V_{cho}$, and
$V_{schm}$ for the model potential of Eq.(24). Two different
strength parameters, $x=0.1$ and $x=0.55$, have been used to see how the 
intruder states 
influence the effective interactions. Our results for these two parameters
are shown in Tables I and II. First let us inspect the Hermiticity of our 
effective interactions; clearly  $V_{okb}$, $V_{cho}$, and $V_{schm}$ are 
all Hermitian, irrespective  
of the strength parameter, as they should be.  The 
degree of non-Hermiticity of $V_{LS}$, however, is highly
dependent on $x$.  In Table I, we see that $V_{LS}$ is only slightly 
non-Hermitian, as the largest difference in symmetric matrix elements 
is only of the order of 20\%.  The impact of the strength parameter on the 
non-Hermiticity
of $V_{LS}$ can be seen when comparing this with table II.  Here $V_{LS}$ 
is strongly non-Hermitian with some symmetric matrix elements differing 
by more than a
factor of 4.  Thus we see that when no intruder states are present (low 
strength parameter), our
parent interaction is approximately Hermitian, but when the intruder states
enter (high strength parameter), our parent interaction, namely $V_{LS}$, 
loses that Hermiticity in a striking manner.

The next point to note concerning Tables I and II is the differences
between the Hermitian effective interactions given by the various
methods.  In Table I, we see that the 
Hermitization procedures produce potentials which do not differ greatly
from the parent potential. This, however, is  expected since 
the parent potential is already approximately Hermitian.  For a high 
strength parameter, we see that the resultant Hermitized potentials 
($V_{okb},V_{cho},V_{schm}$) are
all indeed quite different. Thus, if we want Hermitian potentials which are
similar, it is crucial that the influence of intruder states be minimal.

\begin{figure}
\rotatebox{270}{\scalebox{0.4}{\includegraphics{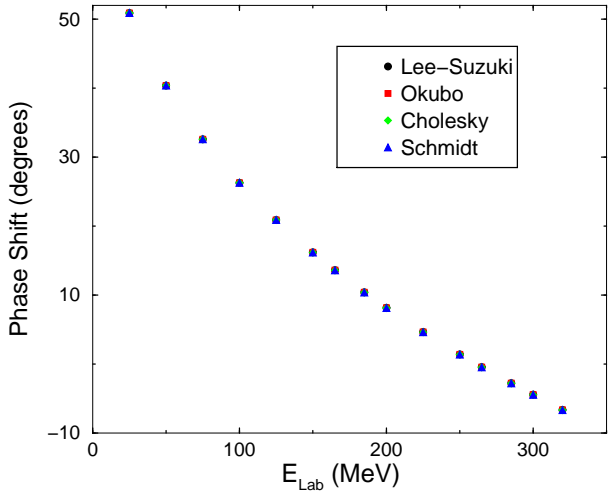}}}
\caption{The phase shifts obtained from the non-Hermitian L-S $V_{low-k}$
are compared with those obtained from the three Hermitian $V_{low-k}$ 
interactions in the $^1S_0$ partial wave channel.}
\label{fig.2}
\end{figure}

Finally, we examine Table III where we show the eigenenergies and 
wavefunctions for the parent and Hermitized potentials. As expected, we see
that the Hermitization procedures preserve eigenenergies.
Note, however, the eigenfunctions of the various interactions
are very different, although they correspond to the same eigenvalues.
We note also that the ground state wave function of both $V_{LS}$
and $V_{schm}$ are equal to the P-space projection of the
ground state wave function of the full-space Hamiltonian.

\section{Phase shift equivalence}

 The non-Hermitian $V_{LS}$ given by the Lee-Suzuki (or folded
diagram) method is specifically constructed to preserve the
half-on-shell T-matrix $T(p',p,p^2)$ \cite{bogner01,bogner02}
; this interaction of course preserves
the phase shift which is given by the fully-on-shell
T-matrix $T(p,p,p^2)$. It would be of interest to study if phase shifts
are also generally preserved by the Hermitian interactions
generated using the transformations described in Section II.

 Let us consider two T-matrices
$T_1(\omega _1)=V_1+V_1g_1(\omega _1)T_1(\omega _1)$ and
$T_2(\omega _2)=V_2+V_2g_2(\omega _2)T_2(\omega _2)$ 
, with the propagators
 $g_1(\omega _1)=\frac{P}{\omega_1-H_0}$ and
 similarly for $g_2(\omega _2)$. The unperturbed state is defined by $H_0$,
namely $H_0 |q\rangle =q^2 |q \rangle$. The symbol $P$ denotes the
principal value boundary condition. These T-matrices are related by the
well known two-potential formula
\begin{equation}
T_2^{\dagger}=T_1+T_2^{\dagger}(g_2-g_1)T_1
+\Omega_2^{\dagger}(V_2-V_1)\Omega _1,
\end{equation}
where the wave operator $\Omega$ is defined by $T_1(\omega _1)
=V_1\Omega_1(\omega _1)$
and similarly for $\Omega _2$.
 Applying the above relation to the half-on-shell T-matrices
in momentum space, we have
\begin{eqnarray}
\langle p' | T_2^{\dagger}(p'^2) |p \rangle
&=& \langle p' | T_1(p^2) |p\rangle \nonumber \\
&& +\langle p' | T_2^{\dagger}(p'^2)(g_2(p'^2)-g_1(p^2))T_1(p^2) |p\rangle
\nonumber \\
&& +\langle \psi_2(p') | (V_2-V_1) |\psi _1 (p) \rangle.
\end{eqnarray}
 Here the true and unperturbed wave
functions are related by $|\psi_1 (p)\rangle =\Omega_1(p^2) |p \rangle $
and similarly for $\psi _2$.

\begin{figure}
\rotatebox{270}{\scalebox{0.4}{\includegraphics{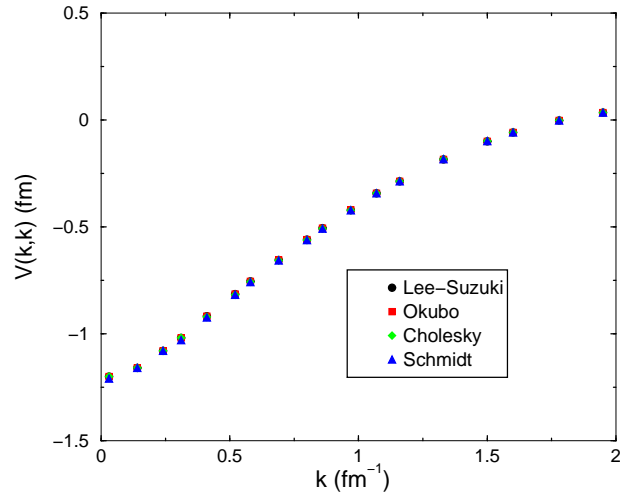}}}
\caption{The non-Hermitian Lee-Suzuki $V_{low-k}$ interaction is compared 
with the three Hermitian $V_{low-k}$ interactions in the $^1S_0$ channel.
}
\label{fig.3}
\end{figure}
 Using the above relation, we shall now show that the phase shifts of
the full-space interaction $V$ are preserved by the Hermitian interaction
$V_{herm}$, for momentum $\leq \Lambda$.
 Let us denote the last term of Eq.(29) as $D(p',p)$.
We use $V_{herm}$ for $V_2$ and  $V$ for $V_1$.
Recall that the eigenfunction of $(H_0+V_{herm})$ is $v_m$ (see Eq.(12))
and that for $H\equiv (H_0+V)$ is $\Psi _m$. We define a wave operator
\begin{equation}
U_P=\sum _{m \in P} |v_m \rangle \langle \Psi_m |.
\end{equation}
Then  $|v_m \rangle =U_P |\Psi _m \rangle$  and 
$PV_{herm}P=U_P(H_0+V)U_P^{\dagger}-PH_0P$. 
 For our present case, $\langle \psi _2(p')|$
is $\langle v_{p'}|$ and $|\psi _1(p)\rangle$ is $|\Psi _p\rangle$. 
Since $\langle v_{p'}|U_P =\langle \Psi _{p'} |$, we have
\begin{eqnarray}
D(p',p)&=&
\langle \Psi_{p'} | (HU_P^{\dagger}-U_P^{\dagger}H) |\Psi _p \rangle,
\nonumber \\
&=& (p'^2-p^2)\langle v_{p'}|\Psi _p \rangle.
\end{eqnarray}
Clearly $D(p,p)=0$. The second term on the right hand side of Eq.(29)
vanishes when $p'=p$. Hence 
\begin{equation}
\langle p | T_{herm}(p ^2) |p \rangle
=\langle p | T(p ^2) |p \rangle,~p\leq \Lambda,
\end{equation}
where $T_{herm}$ is the T-matrix for $(H_0+V_{herm})$ and 
$T$ for $(H_0+V)$. 
Consequently the phase shifts of  $V$ are 
preserved by $V_{herm}$.
 Recall that our T-matrices are real, because of
the principal-value boundary conditions employed.

\begin{figure}
\rotatebox{270}{\scalebox{0.4}{\includegraphics{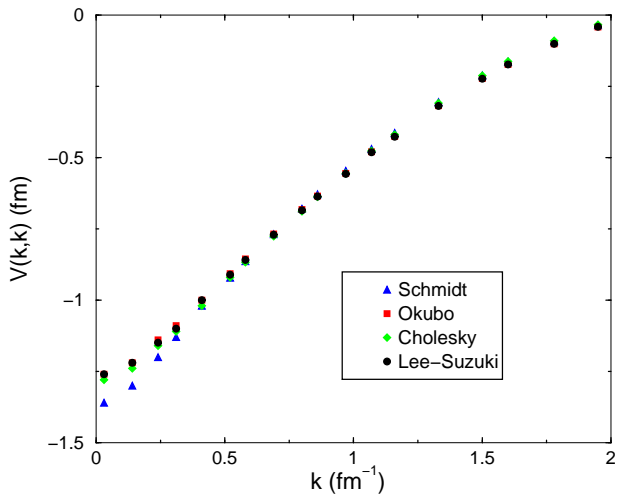}}}
\caption{The non-Hermitian Lee-Suzuki $V_{low-k}$ interaction is compared 
with the three Hermitian $V_{low-k}$ interactions in the $^3S_1$ channel.
}
\label{fig.4}
\end{figure}

To double check this preservation, we have calculated the phase shifts of
the various Hermitian potentials ($V_{okb},V_{cho}$ and $V_{schm}$)
together with  $V_{LS}$. As shown in  Fig. 2 
the $^1S_0$ phase shifts obtained from them 
all agree with each other very well, as is the case for other partial waves.
The calculations were performed with the CD-Bonn potential and
$\Lambda=~2.0fm^{-1}$.
Since the phase shifts of $V_{LS}$
are, by construction, the same as those of the full-space potential $V$,
the Hermitian potentials preserve the phase shifts of $V$.
It has been pointed out \cite{epel99} that the Okubo Hermitian potential
preserves the phase shifts, but we have found that there is a family of
Hermitian potentials, including Okubo, which all preserve
the phase shifts up to the decimation scale $\Lambda$.

\section{Hermitian low-momentum interactions}

A main purpose of having a low-momentum nucleon interaction $V_{low-k}$
is to use it in nuclear many body problems such as the shell model
nuclear structure calculations. As we have seen, however, there is
a family of phase-shift equivalent Hermitian interactions.
How different are they?
Which one should one use for nuclear structure calculations?
Will these Hermitian effective interactions give rise to the same 
physical properties? It is these questions we now seek to answer.


We have calculated the Hermitian $V_{low-k}$'s corresponding to
$V_{okb}$, $V_{cho}$ and $V_{schm}$ using several NN potentials.
Calculations for the non-Hermitian $V_{low-k}$ corresponding to
$V_{LS}$ were also performed. In Fig. 3 we compare the results
for the $^1S_0$ channel, obtained with the CD-Bonn potential
and $\Lambda=~2.0fm^{-1}$.  Clearly the
Hermitian interactions are all quite similar to each other and to the 
parent non-Hermitian potential $V_{LS}$. In Fig. 4, we show a 
similar plot for the $^3S_1$ channel. 
Again, with the exception of very low momentum, 
 the Hermitian potentials are all nearly identical to the parent $V_{LS}$.
It is of interest that at very low momentum, the $V_{schm}$ matrix elements
are slightly more attractive than the others. 
We note that the Hermitian effective interactions
all preserve the deuteron binding energy (2.225 MeV). 
 In addition, they are all phase shift equivalent
up to the decimation scale $\Lambda$, as illustrated in Fig. 2.

\begin{figure}
\rotatebox{270}{\scalebox{0.4}{\includegraphics{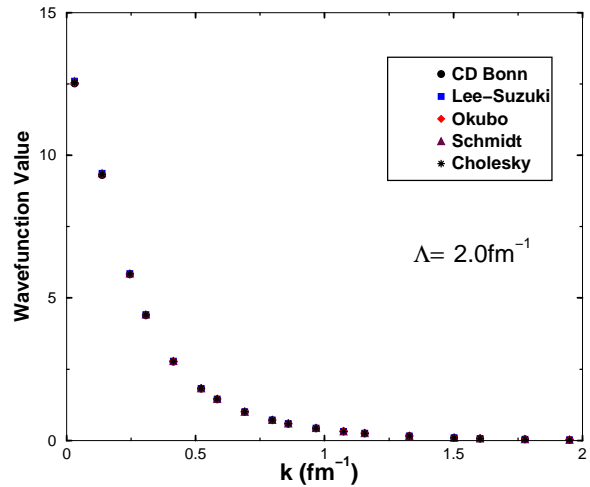}}}
\caption{S-state deuteron wavefunctions for the bare and effective 
interactions plotted with respect to momentum, using a model space cutoff of
$\Lambda = \rm 2.0 fm^{-1}$.}
\label{fig.5}
\end{figure}

\begin{table*}
\caption{A comparison of shell model relative matrix element calculated
using $V_{LS}$ with those calculated using the Hermitian interactions.
The oscillator length parameter $b$  is given by 
$b=\sqrt{\frac{\hbar}{m\omega}}$, 
 in units of $fm$. The matrix elements are in units of $MeV$.}
\vskip 0.5cm
\begin{tabular}
{cc|cccc|cccc}
\hline
 & 
\multicolumn{1}{c} {b(fm)} & 
\multicolumn{4}{c} {$n=0$} &
\multicolumn{4}{c} {$n=1$}    \\
\hline
$ \langle n^{1}S_{0} \mid  V  \mid n^{1}S_{0}  \rangle $ & & $V_{LS}$ & 
$V_{okb}$ &$V_{cho}$ & $V_{schm}$ &$V_{LS}$ & $V_{okb}$ &$V_{cho}$ & 
$V_{schm}$ \\
 & 1.4 & -9.96 & -9.96 & -9.96 & -9.96 & -5.43 & -5.43 & -5.43 & -5.43 \\
 & 2.0 & -4.85 & -4.85 & -4.85 & -4.85 & -4.40 & -4.40 & -4.40 & -4.40 \\   
 & 2.6 & -2.59 & -2.59 & -2.60 & -2.60 & -2.80 & -2.80 & -2.81 & -2.81 \\ 

$ \langle n^{1}S_{0} \mid  V  \mid (n+1)^{1}S_{0}  \rangle $& & & & & & & 
& &\\
 & 1.4 & -6.98 & -6.98 & -6.98 & -6.98 & -3.06 & -3.06 & -3.06 & -3.06 \\
 & 2.0 & -4.48 & -4.48 & -4.48 & -4.48 & -3.83 & -3.83 & -3.83 & -3.83 \\
 & 2.6 & -2.64 & -2.64 & -2.65 & -2.65 & -2.69 & -2.69 & -2.70 & -2.70 \\ 

$ \langle n^{1}S_{0} \mid  V  \mid (n+2)^{1}S_{0}  \rangle $& & & & & & & 
& &\\
 & 1.4 & -4.04 & -4.04 & -4.04 & -4.04 & -1.06 & -1.06 & -1.06 & -1.06 \\
 & 2.0 & -3.80 & -3.80 & -3.80 & -3.80 & -3.13 & -3.13 & -3.13 & -3.13 \\
 & 2.6 & -2.49 & -2.49 & -2.49 & -2.49 & -2.49 & -2.49 & -2.49 & -2.49 \\

$ \langle n^{3}S_{1} \mid  V  \mid n^{3}S_{1}  \rangle $    & & & & & & & 
& &\\
 & 1.4 & -12.29& -12.29& -12.29& -12.29& -9.11 & -9.11 & -9.11 & -9.11 \\
 & 2.0 & -5.51 & -5.51 & -5.51 & -5.51 & -5.70 & -5.70 & -5.70 & -5.70 \\
 & 2.6 & -2.85 & -2.84 & -2.85 & -2.85 & -3.32 & -3.31 & -3.33 & -3.33 \\

$ \langle n^{3}S_{1} \mid  V  \mid (n+1)^{3}S_{1}  \rangle $& & & & & & & 
& &\\
 & 1.4 & -10.18& -10.18& -10.18& -10.18& -6.65 & -6.65 & -6.65 & -6.65 \\ 
 & 2.0 & -5.46 & -5.46 & -5.46 & -5.46 & -5.33 & -5.33 & -5.33 & -5.33 \\ 
 & 2.6 & -3.02 & -3.01 & -3.02 & -3.02 & -3.32 & -3.31 & -3.30 & -3.32 \\
\end{tabular}
\end{table*}

Despite the similarities, there is, however, an additional degree of 
preservation that
$V_{schm}$ satisfies but the other two Hermitian interactions don't. 
By construction, the deuteron ground-state wave function given by 
$V_{schm}$ is exactly equal
to the P-space projection of the wave function of $V$, which is not true
for $V_{okb}$ and $V_{cho}$. This additional preservation is worth studying
further. We refer to Figures 5 and 6, where we
plot the S and D-state deuteron wavefunctions for a cutoff of 
$\Lambda= \rm 2.0  fm^{-1}$. Overall, the agreement is very good between all
potentials, not just $V_{LS}$ and $V_{schm}$, which is to be expected 
considering that the interactions themselves are approximately the same.
The fact that $V_{schm}$ gives 
exactly the same wavefunction as $V_{LS}$ can be seen from the D-state
probability of the deuteron for each interaction, which we list
for convenience in Fig. 6.
Whereas the $P_D$'s for the effective interactions are close, they are exactly equal
for $V_{LS}$ and $V_{schm}$. This exact preservation of the ground state 
wavefunction presents us with an extra constraint that might be useful in 
deciding which Hermitian potential to use.

To ensure our Hermitian potentials will be useful for nuclear structure
calculations, 
we examine in Table IV some shell model matrix elements calculated
with $V_{low-k}$ corresponding to $V_{okb}$, $V_{cho}$, $V_{schm}$
and $V_{LS}$.
As seen they are all virtually identical, as would be expected
owing to the similarity between the potentials themselves.
This good agreement is a desirable result, as it implies that
shell model calculations will not be sensitive to which
interaction one employs.

We have seen that the Hermitian potentials generated above are all 
approximately the same, and we would like to offer an explanation as to why 
this is so. Although the
$V_{low-k}$ corresponding to $V_{LS}$ is non-Hermitian, 
we emphasize that it is only slightly so. 
In reference to our model study of Section 2, this corresponds 
to the situation  with a small strength 
parameter, and thus it would not be surprising to see the Hermitian 
$V_{low-k}$ potentials are so similar to the parent $V_{low-k}$. For example,
this is
especially transparent in the case of the Okubo interaction of Eq. (18) 
or (19), as it can be
written as \cite{suzuki83,kuo93}
\begin{eqnarray}
\lefteqn{\langle \alpha |V_{okb}| \beta \rangle =
\langle \alpha | \frac{1}{2}(V_{LS}+V_{LS}^{\dagger})|\beta\rangle}  \\
& &  +\frac{\sqrt{\mu_{\alpha}^2+1}-\sqrt{\mu_{\beta}^2+1}}
{\sqrt{\mu_{\alpha}^2+1}+\sqrt{\mu_{\beta}^2+1}}
\langle\alpha|\frac{1}{2}(V_{LS}-V_{LS}^{\dagger})|\beta\rangle\ . \nonumber 
\end{eqnarray}
This tells us that $V_{okb}$ can be well approximated by 
 $V_{LS}$, if $V_{LS}$ is only slightly non-Hermitian. 
In fact in this case $V_{okb}$ is very accurately
reproduced by the simple average $(V_{LS}+V_{LS}^{\dagger})/2$.

We note that the D-state 
probabilities given in Fig.6 
 for the various effective interactions are 
significantly different from that of the bare potential.  This is
due to the fact that the effective interactions are subjected to a momentum 
cutoff at $\rm 2.0~ fm^{-1}$, while the bare CD Bonn potential extends far 
beyond that. This can clearly be seen from the figure where a large portion
of the wavefunction is simply cut off. While $P_D$ is not an observable, it 
has theoretical relevance-it is an important characteristic of
modern nucleon-nucleon potentials.

According to the MEEFT prescription, it is necessary to impose the cutoff at
the limit of experimental data (in this case the limit of NN scattering 
experiments at $\rm 2.0~ fm^{-1}$), and it is at this cutoff that the effective
interactions from all the bare potentials become nearly identical.
 Raising
the cutoff would erode this approximate uniqueness and model independence. 
But at the above  cutoff, we can not preserve $P_D$.  To
illustrate, we refer to Figure 7 where we plot the D-state wavefunctions 
with a cutoff of $\Lambda= \rm 3.0~ fm^{-1}$. It shows that much more of the
wavefunction is contained in the region below the cutoff, and as a result,
the $P_D$'s for the effective interactions are much closer to that of the 
bare potential. If we increase the cutoff to $\rm 4.0 fm^{-1}$, they are almost
exact. How to resolve this disparity in how to choose the cutoff warrants
further study.

\begin{figure}
\rotatebox{270}{\scalebox{0.4}{\includegraphics{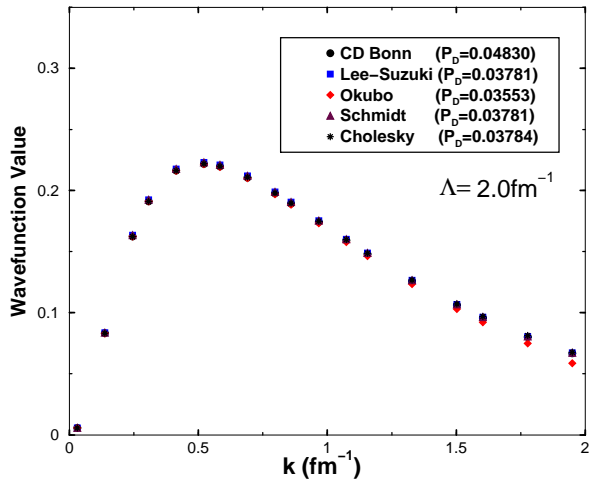}}}
\caption{D-state deuteron wavefunctions for the bare and effective 
interactions plotted with respect to momentum, using a model space cutoff of
$\Lambda = \rm 2.0 fm^{-1}$.}
\label{fig.6}
\end{figure}

\section{Summary and conclusion}
We have studied a general method for deriving low-momentum NN
interactions which are phase shift equivalent.
By integrating out the high momentum components of a realistic 
NN potential such as the CD-Bonn potential,
the Lee-Suzuki or folded-diagram method is employed to derive
 a parent low-momentum NN potential. 
This potential preserves the
deuteron binding energy and phase shifts up to  cutoff scale $\Lambda$.
In addition it preserves the half-on-shell T-matrix up to the same scale.
This Lee-Suzuki low-momentum  
NN interaction is not Hermitian, and further transformation is needed
to obtain low-momentum interactions which are Hermitian.
We have shown how to construct a family of such an interaction using the 
Schmidt orthogonalization procedure, and we have seen that two existing 
Hermitization schemes, namely Okubo and Andreozzi, are in fact special 
cases of our general process. We have shown that all the Hermitian
interactions so generated are phase shift equivalent, all reproducing
empirical phase shifts up to scale $\Lambda$. These potentials
also preserve the deuteron binding energy.

\begin{figure}
\rotatebox{270}{\scalebox{0.4}{\includegraphics{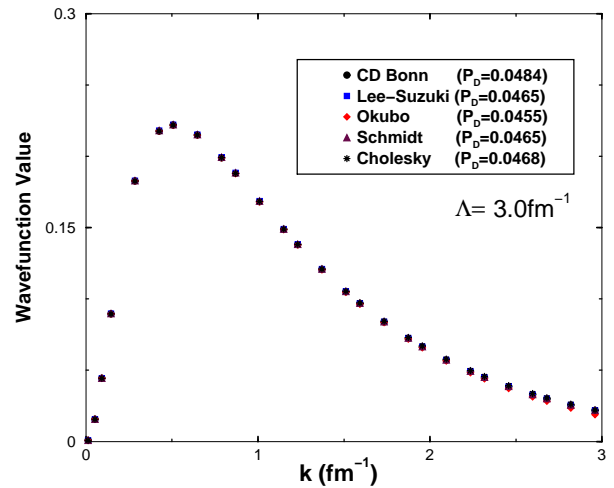}}}
\caption{D-state deuteron wavefunctions for the bare and effective 
interactions plotted with respect to momentum, using a model space cutoff of
$\Lambda = \rm 3.0 fm^{-1}$. } 
\label{fig.7}
\end{figure}

Through an analysis of this procedure using a solvable matrix model  we have 
seen some interesting properties of Schmidt transformation method. 
 In particular, with the 
entrance of intruder states, the parent potential can
become highly non-Hermitian, and that the Hermitian potentials can
deviate largely   from each other and from the  parent potential.  
It is fortunate that such deviations are not present
for low-momentum NN potentials, for cutoff momentum $\Lambda\sim2.0~fm^{-1}$.
This is mainly because our parent $V_{low-k}$
is only slightly non-Hermitian, and as a result, the Hermitian
low-momentum nucleon interactions generated from our orthogonalization
procedure are all close to each other and close to the parent
$V_{low-k}$. Shell model matrix elements of the low-momentum
nucleon interactions are found to be approximately independent of the starting
Hermitian potential, indicating the usefulness of our procedure in nuclear
many body calculations.


\begin{acknowledgments}
 We thank Jeremy Holt for many helpful discussions.
Partial support from the US Department of Energy under contracts
DE-FG02-88ER40388 is gratefully acknowledged.
\end{acknowledgments}

\end{document}